\documentclass[9pt, sigconf]{IEEEtran}
\usepackage{amsmath}
\usepackage{hhline}
\usepackage{silence}
\usepackage{graphicx}
\usepackage{multirow} 
\usepackage{graphicx}
\usepackage{multirow}
\usepackage{amsmath}
\usepackage[english]{babel}
\usepackage{booktabs}
\usepackage{array}
\usepackage{paralist}
\usepackage{threeparttable}
\usepackage{lipsum}
\usepackage{flushend}
\usepackage{cuted}
\usepackage{algpseudocode}
\usepackage{algorithm}
\usepackage{amssymb}
\usepackage{color}
\usepackage{psfrag}
\usepackage{epsfig}
\usepackage{multicol}
\usepackage{color}
\usepackage{fancybox}
\usepackage{subfigure}
\usepackage{cite}
\usepackage{theorem}
\usepackage{url}
\usepackage[table]{xcolor}
\usepackage{epstopdf}
\usepackage[normalem]{ulem}
\usepackage{siunitx}
\usepackage[justification=centering,font=small,labelfont=bf]{caption}
\definecolor{gray1}{gray}{0.90}
\definecolor{gray2}{gray}{0.98}
\definecolor{light-gray}{gray}{0.95}

\newcommand{\ignore}[1]{}
\newcommand{\redHL}[1]{\textcolor{red}{#1}}

\renewcommand{\baselinestretch}{0.97}

\begin{document}

\title{A New, Computationally Efficient ``Blech Criterion'' for Immortality in
General Interconnects}

\author{Mohammad Abdullah Al Shohel, Vidya A. Chhabria, and Sachin S. Sapatnekar\\
Department of Electrical and Computer Engineering\\
University of Minnesota, Minneapolis, MN 55455, USA.\thanks{\scriptsize
This work was supported in part by the NSF under award CCF-1714805, by
the DARPA OpenROAD project, and the Louise Dosdall Fellowship.\newline
978-1-6654-3274-0/21/\$31.00 \copyright 2021 IEEE}
}

\maketitle

\begin{abstract}
Traditional methodologies for analyzing electromigration (EM) in VLSI circuits
first filter immortal wires using Blech's criterion, and then perform detailed
EM analysis on the remaining wires.  However, Blech's criterion was designed
for two-terminal wires and does not extend to general structures.  This paper
demonstrates a first-principles-based solution technique for determining the
steady-state stress at all the nodes of a general interconnect structure, and
develops an immortality test whose complexity is linear in the number of edges
of an interconnect structure.  The proposed model is applied to a variety of
structures.  The method is shown to match well with results from numerical
solvers, to be scalable to large structures.
\end{abstract}

\section{Introduction}
\label{sec:intro}

\noindent
Electromigration (EM) aging in metal wires is caused by material transport of
atoms, triggered by electron current through the wires. EM has become a major
concern in electronic circuits due to the increase in current density.
Previously, EM was considered a problem only in upper metal layers that carry
the largest current, but with scaling, as transistors drive increasing amounts
of current through narrow wires, EM hotspots have emerged through the stack.

The conventional method for EM analysis for interconnects involves a two-stage
process. In the first stage, EM-immune wires are filtered out using the Blech
criterion~\cite{blech:76}, which compares the product of the current density $j$
through a wire with its length, $l$.   This $jl$ product is compared against a
technology-specific threshold, and any wires that fall below this product are
deemed immortal, while others are potentially mortal.  In the second stage, 
wires in the latter class undergo further analysis to check whether or not the
EM failure may occur during the product lifespan.  Traditionally, this involves
a comparison of the current density through these wires against a global limit,
set by the semi-empirical Black's equation~\cite{black:69}; more recent 
approaches include~\cite{Chen2016,Chatterjee18,vivek:dac,Mishra16}.

However, this approach is predicated on analyses/characterizations of
single-wire-segment test structures, which determine the critical $jl$ product
threshold for the Blech criterion, and the upper bound on $j$ in Black's
equation. In practice, wires typically have multiple segments with
different current densities. The criterion for immortality under this scenario
is quite different from the Blech criterion, and while the limitations of the
criterion have been widely recognized in past work, there is no computationally
simple test similar to the Blech criterion to determine immortality for general
interconnects.

As opposed to the empirical Black's equation based approach, there has been an
emerging thread on using physics-based analysis for EM in interconnects.
Building upon past work such as~\cite{rosen:71,Schatzkes86,Clement92}, the work
in~\cite{kor:93} presented a canonical treatment of EM equations in a metallic
interconnect, with exact solutions for a semi-infinite and finite line.  This
paper has formed the basis of much work since then, with techiques that attempt
to obtain solutions for a single-segment lines~\cite{Sukharev14,Chen2016}.  For
multisegment lines, several attempts have been made to solve the general
transient analysis problem~\cite{Chen2016,Chatterjee18} through detailed
simulations, but the key to checking for immortality is to solve the {\em
steady-state} problem.  The methods in~\cite{Riege98,Clement99}, subsequently
extended in~\cite{ala:05}, used a sum of $jl$ products along wire segments: if
$j_i$ is the current density through the $i^{\rm th}$ segment of length $l_i$,
then the largest $\sum j_i l_i$ on any path in a tree was taken to be the
worst-case stress: as observed in~\cite{Abbasinasab15}, this is incorrect. 

\ignore{An experimentally-driven method in~\cite{parkvianode:10} observed an
apparently counterintuitive observation in multisegment wires: that failures
can occur sooner in segments with lower current density. It presented a
heuristic approach for finding an effective current density. As we will show,
this scenario can be explained using our approach, and a more precise
formulation for the effective current density can be determined.}
In~\cite{Haznedar06}, a system of equations describing steady-state analysis in
an interconnect tree was presented and solved. However, the structure of the
difference equations was not exploited to obtain a generalizable solution.  The
analyses in~\cite{LinOates13,Lin16} solve a related problem for a simple two-
or three-segment structure with a passive reservoir.  The work in~\cite{Sun18}
develops analysis principles and applies them to several structures, with
closed-form formulas for simple topologies.  However, it does not provide a
scalable algorithm for general structures.


Thus, there is no truly general, scalable formula for immortality detection
to replace the Blech criterion.  This work solves this problem with a
linear-time algorithm for general multisegment interconnects.  On
comparable CPUs, our approach provides solutions to IBM PG benchmarks in a
few minutes, while~\cite{Sun18} requires over an hour.
\ignore{
The focus of this work is purely on
determining a linear-time technique to replace the Blech criterion to filter
wires for immortality.  Once this updated criterion is applied to filter out
immortal wires, existing
methods~\cite{Chen2016,Chen2017,Chatterjee18,vivek:dac,Mishra16,Li11}, can be
applied to the remaining, potentially mortal, wires to determine whether they
fail during the chip lifetime.
}

\section{Background}

\begin{figure}
\centering
\includegraphics[width=0.7\linewidth]{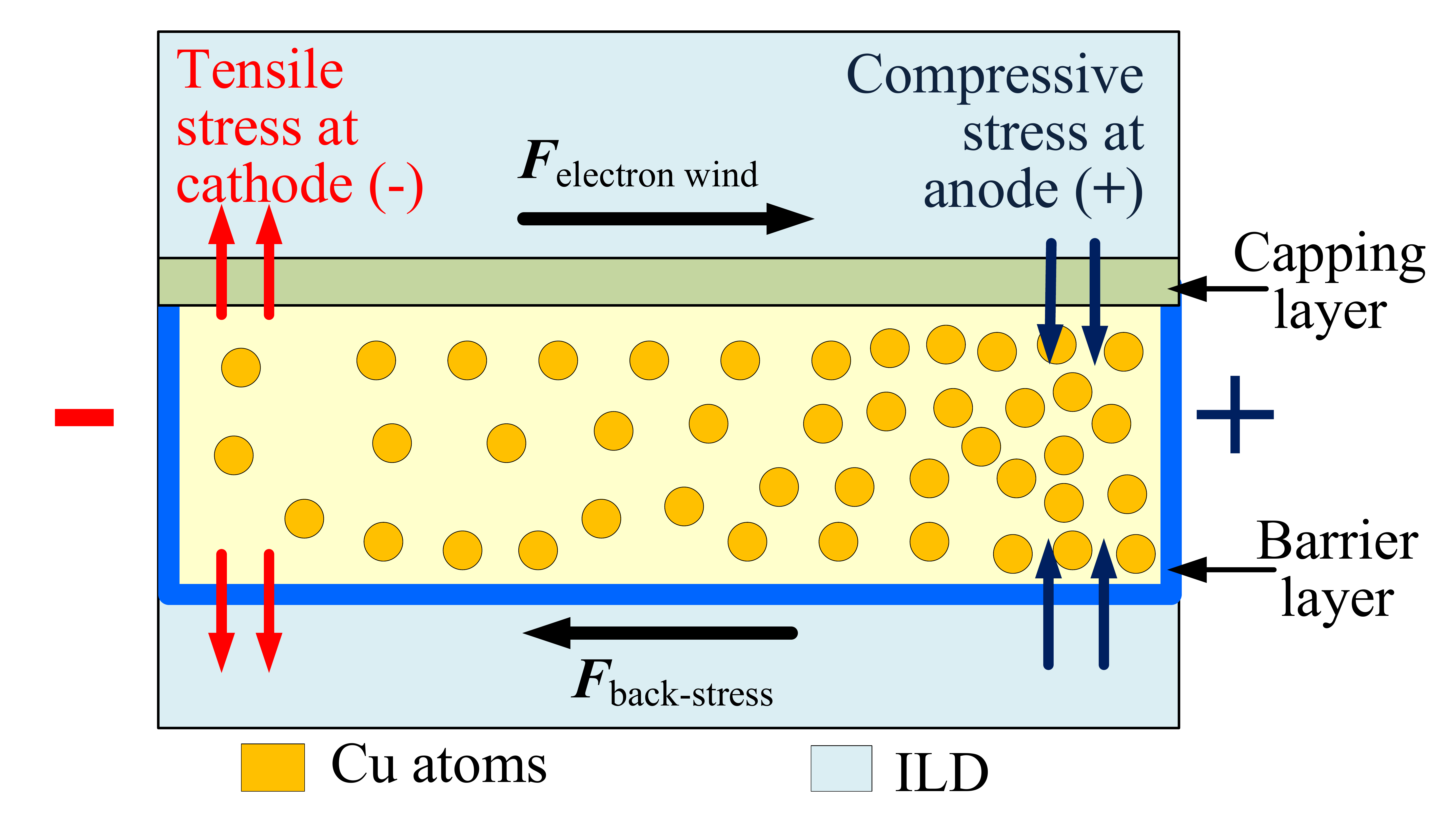}
\caption{Cross section of a Cu wire indicating the electron wind force and
back-stress force~\cite{Mishra16}.}
\label{fig:Cu DD wire}
\vspace{-4mm}
\end{figure}

\noindent
Figure \ref{fig:Cu DD wire} illustrates the electromigration mechanism in a Cu
dual-damascene (DD) wire. 
As the current flows in a metal wire, metal atoms are transported from the
cathode towards the anode, in the direction of electron flow, by the momentum
of the electrons.  This electron wind force causes a depletion of metal atoms
at the cathode, potentially resulting in void formation, leading to open
circuits.  In a Cu DD interconnect, the movement of migrating atoms occurs in a
single metal layer, and atoms are prevented from
migrating to other metal layers due to the capping or barrier layer, which acts
as a blocking boundary for mass transport~\cite{Gambino18,Zhang10}.
Consequently, within a metal layer, mass depletion of atoms occurs at the
cathode terminal and mass accumulation occurs at the anode terminal.  A tensile
stress is created near the anode, and a compressive stress near the cathode. 

The concentration gradient caused by metal migration creates a tendency for
atoms to diffuse back to the cathode.  This force, acting against the electron
wind, is proportional to the stress gradient. 

\subsection{Notation}

\noindent
For a general interconnect structure with multiple segments, we define the
following notation. This is represented by an undirected graph ${\cal
G}(V,E)$ with $|E|$ segments and $|V|$ nodes.  The vertices $V = \{ v_1,
\cdots, v_{|V|} \}$ are the set of {\em nodes} in the structure, and edges $E =
\{ e_1, \cdots, e_{|E|} \}$ are the set of wire {\em segments}.  A vertex of
degree 1 is referred to as a {\em terminus}.

\begin{figure}
\centering
\includegraphics[width=0.95\linewidth]{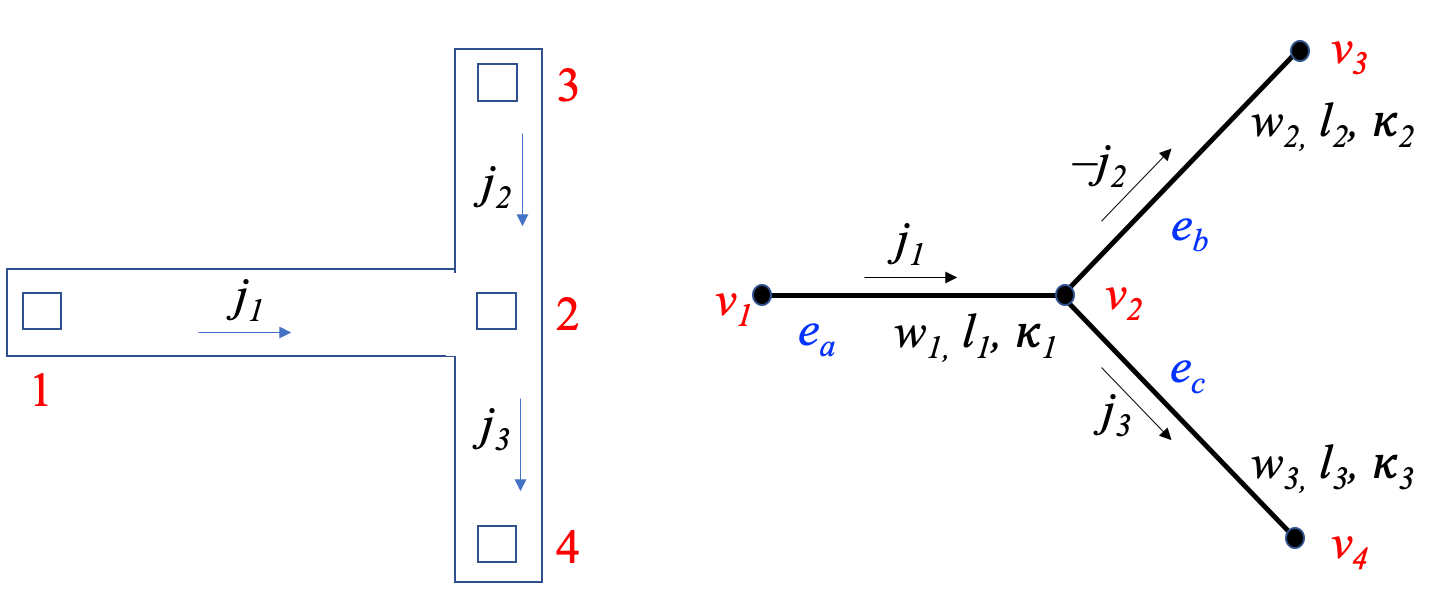}
\caption{(a) A simple net fragment. (b) Its equivalent graph, with arrows showing
the reference current direction for each edge.}
\label{fig:netfragment}
\vspace{-4mm}
\end{figure}

Each edge $e_i$ is associated with a reference current direction, and has three
attributes: length $l_i$, width $w_i$, and current density $j_i$.  The sign of
current density is relative to the reference direction of the edge: it is
negative if the current is opposite to the reference direction.
Figure~\ref{fig:netfragment} shows a net fragment and its graph model for a
tree with four nodes and three edges: since the current direction in $e_b$ is
opposite to the reference direction, the current density is shown as $-j_2$.

Along each edge, we use a {\em local coordinate system} along each segment $i$.
If the edge has a reference direction from node $v_a$ to node $v_b$, we
represent the position of node $a$ as $x=0$ and that of node $b$ as $x=l_i$.
As part of our analysis, we will compute stresses induced within the
interconnect.  Specifically,
\begin{itemize}
\item $\sigma_i(x,t)$ is the stress within wire {\em segment} $i$ at time
$t$ at a location $x$, where $0 \leq x \leq l_i$ and $1 \leq i \leq |E|$.
\item $\sigma^k$ is the steady-state stress at {\em node} $v_k$,
$1 \leq k \leq |V|$.
\end{itemize}

\subsection{Stress equations for interconnect structures}
\label{sec:stress_structures}

\noindent
A single interconnect segment injects electron current at a cathode at $x=0$
towards an anode at $x=l_i$.  The temporal evolution of EM-induced stress,
$\sigma(x,t)$, in the segment is modeled as~\cite{kor:93}:
\begin{align}
\frac{\partial \sigma}{\partial t} &= 
  \frac{\partial }{\partial x} \left [
         \kappa \left ( \frac{\partial \sigma}{\partial x} + \beta j_i
                \right) \right ]
\label{eq:Korhonen's_eqn}
\end{align}
Here, $x$ is the distance from the cathode; $\beta = (Z^* e \rho)/\Omega$;
$j_i$ is the current density through the wire; $Z^*$ is the effective charge 
number; $e$ is the electron charge; $\rho$ is the resistivity; and $\Omega$ is the atomic volume for the
metal (in the literature, $\beta j_i$ is often denoted as $G$).  Here, $\kappa
= D_a {\cal B} \Omega/(kT)$, where ${\cal B}$ is the bulk modulus of the
material, $k$ is Boltzmann's constant, and $T$ is the temperature, $D_a=D_0
e^{-E_a/kT}$ is the diffusion coefficient, with $E_a$ being the activation
energy.  The boundary conditions (BCs) depend on wire topology.

When no current is applied, the stress in the wire is given by $\sigma_T$, the
thermally-induced stress due to differentials in the coefficient of thermal
expansion (CTE) in the materials that make up the interconnect stack.  The
differential equation with the boundary conditions can be solved numerically to
obtain the transient behavior of stress over time.  Due to superposition, the
stress in the wire can be computed in this way and $\sigma_T$ can then be added
to account for CTE effects.  The impact of $\sigma_T$ is to offset the critical
stress, $\sigma_{crit}$, to $(\sigma_{crit} - \sigma_T)$.

As in~\cite{kor:93}, the sign convention for $j_i$ is in the direction of
electron current, i.e., opposite to conventional current and the electric
field.  The atomic flux attributable to the electron wind force is proportional
to the second term on the right hand side that contains $j_i$, while the flux
related to the back-stress force is proportional to the first term containing
the stress gradient $\frac{\partial \sigma}{\partial x}$.  In both cases, the
constant of proportionality varies linearly with the cross-sectional area of
the wire.  The sum, $(\partial \sigma/\partial x + \beta j_i)$, is proportional
to the net atomic flux. 

\noindent
{\bf BCs for single-segment interconnect}
When electron current is injected through the anode and flows to
the cathode at the other end, we have zero-flux conditions at each end:
\begin{align}
\frac{\partial \sigma}{\partial x} + \beta j_1 = 0 \; \;
		\forall \; t \mbox{ at $x=0, x=l_1$.}
\label{eq:BC_single}
\end{align}

\noindent
{\bf BCs for a multisegment interconnect trees/meshes}
The boundary conditions at the terminus nodes (i.e., nodes of degree 1) 
require zero flux across the blocking boundary, i.e.,
\begin{align}
\left . \frac{\partial \sigma_e}{\partial x} 
           \right |_{\mbox{\footnotesize{terminus}}} + \beta j_e  = 0
\label{eq:BC_terminal_tree}
\end{align}
where edge $e$ connected to the terminus has current density $j_e$.

For any internal node $n$ of the structure with degree $d \geq 2$, let
the incident edges with reference current directed into the
node be $\{e_1, \dots , e_m\}$, and the edges directed away from the node be
$\{e_{m+1}, \dots , e_d\}$; if either set is empty, $m=0$ or $d$.  The flux
boundary conditions at such a node are given by
\begin{align}
\sum_{k \in \{1, \cdots , m \}} w_{e_k} 
    &\left ( \left . \frac{\partial \sigma_{e_k}}{\partial x} 
            \right |_n + \beta j_{e_k}
    \right ) = 
\label{eq:BC_internal_tree_flux} \\
& \sum_{k \in \{m+1, \cdots , d \}} w_{e_k} 
    \left ( \left . \frac{\partial \sigma_{e_k}}{\partial x} 
            \right |_n + \beta j_{e_k}
    \right )
\nonumber
\end{align}
and the continuity boundary conditions are:
\begin{align}
& \sigma_{e_1} |_n = \sigma_{e_2} |_n = \cdots = \sigma_{e_d} |_n
\label{eq:BC_continuity_tree} 
\end{align}
where $\sigma_{e_k} |_n$ and $\partial \sigma_{e_k}/\partial x |_n$ are the
values of the stress and its derivative at the location corresponding to node
$n$.

\section{Analysis of Steady-state Stress}

\subsection{Equations for steady-state analysis in a wire segment}
\label{sec:segment_solution}

\noindent
We will work with \eqref{eq:Korhonen's_eqn} as a general representation
of the stress in any multisegment line or tree.  In the steady state, when the
electron wind and back-stress forces reach equilibrium, then for each segment
$i$, over its entire length, $0 \leq x \leq l_i$,
\begin{align}
& \frac{\partial \sigma_i}{\partial x} + \beta j_i  = 0,
\mbox{  i.e.,  } \frac{\partial \sigma_i}{\partial x} = -\beta j_i
\label{eq:constant_slope}
\end{align}
%
The Blech criterion for immortality in a single-segment line states that in the
steady state, if the maximum stress falls below the critical stress,
$\sigma_{crit}$, required to nucleate a void, then the wire is considered
immortal, i.e., immune to EM.  This translates to the condition~\cite{blech:76}:
\begin{align}
j l \leq  (jl)_{crit} 
\label{eqn:Blech_criterion}
\end{align}
where $(jl)_{crit}$ is a function of the critical stress, $\sigma_{crit}$. 

The derivation of the Blech criterion is predicated on the presence of blocking
boundary conditions at either end of a segment carrying constant current, and
is invalid for multisegment wires, even though it has been (mis)used in that
context.  For a general multisegment structure, from \eqref{eq:constant_slope},
{\em a linear gradient exists along each segment} of a general multisegment
structure (this has been observed for multi-segment
lines~\cite{Riege98,Clement99} and meshes~\cite{Haznedar06}).  

\noindent
{\em Lemma~1}: 
For edge $e_k$ with reference current direction from vertex $v_a$ to
$v_b$, the steady-state stress along the segment is:
\begin{align}
\sigma_k(x) &= \sigma^a -\beta j_k x 
\label{eq:linearstress1} \\
\mbox{ and }
\sigma^b - \sigma^a &= -\beta j_k l_k
\label{eq:linearstress2}
\end{align}
where $\sigma^a$ ($\sigma^b$) denotes the steady-state stress at node $a$ ($b$).

\noindent
{\em Proof:}
The first expression follows directly from~\eqref{eq:constant_slope}, and the
second is obtained by substituting $x=l_k$ at node $v_b$.
\hfill $\Box$

The following corollary follows directly from~\eqref{eq:linearstress1}:\\
{\em Corollary 1}:
For edge $e_k = (v_a,v_b)$ in an interconnect structure, 
\begin{align}
\int_0^{l_k} \sigma_k(x) dx =
\int_0^{l_k} (\sigma^a - \beta j_k x) dx =
\sigma^a l_k - \beta j_k \frac{l_k^2}{2}
\label{eq:segment_integral}
\end{align}

\noindent
{\em Corollary 2}: In a segment, the largest stress is at an end point.

\noindent
{\em Proof:}
This follows from~\eqref{eq:linearstress2}: if $j_k \geq 0$, the stress on the
segment is maximized at node $v_a$; otherwise at node $v_b$.
\hfill $\Box$

\subsection{Equations for steady-state analysis in a general structure}
\label{sec:general_solution}

\noindent
The existence of cycles in a graph requires careful consideration: we show
that the solution can be found by analyzing a spanning tree.

\begin{figure}[htb]
\centering
\includegraphics[width=0.33\linewidth]{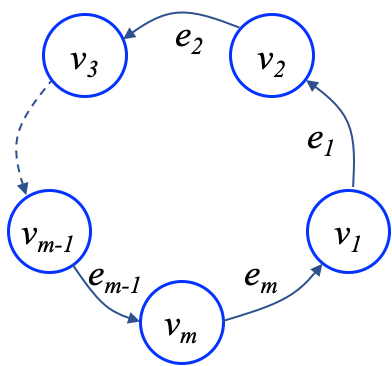}
\caption{A cycle in ${\cal G}(V,E)$.}
\label{fig:cycle}
\vspace{-2mm}
\end{figure}

\noindent
{\em Theorem 1:} 
Consider any undirected simple cycle, without repeated vertices or edges, $\cal
C$ in ${\cal G}(V,E)$, consisting of edges $e_1, \cdots, e_m$ containing
vertices $v_1, v_2, \cdots, v_m$, with edge reference directions from $v_i$ to
$v_{i+1}$ (where $v_{m+1} \stackrel{\Delta}{=} v_1$), as shown in
Fig.~\ref{fig:cycle}.  The $m$ steady-state stress
equations~\eqref{eq:linearstress2} representing this cycle
are linearly dependent.  A linearly independent set of equations
is obtained by dropping one equation, i.e., breaking the cycle by dropping
one edge.

\noindent
{\em Proof:}
Let $V_i$ be the voltage at vertex $v_i$, $R_i$ be the resistance of wire
segment $i$, $\rho$ be the wire resistivity, and $h_i$ be the wire thickness
(constant in layer $i$). Then $R_i = \rho l_i/(w_i h_i)$ and by Ohm's
law,
\begin{align}
j_i = (V_{i+1} - V_{i})/(R_i w_i h_i) = 
(V_{i+1} - V_{i})/(\rho l_i)
\end{align}
According to~\eqref{eq:linearstress2}, along each edge $e_i = (v_i, v_{i+1})$,
\begin{align}
\sigma^{i+1} - \sigma^i &= -\beta j_i l_i = -\beta (V_{i+1} - V_i)/\rho
\label{eq:linearstress3}
\end{align}
Adding up all equations~\eqref{eq:linearstress3} around the cycle, the left
hand side sums up to zero, because each $\sigma^k$ term in one equation has a
corresponding $-\sigma^k$ term in the next equation (modulo $m$, so that
$-\sigma^1$ and $\sigma^1$ appear in the last and first equation,
respectively).  Similarly, the right-hand side also sums up to zero due to
telescopic cancelations of $V^k$ in each equation and $-V^k$ in the next
equation (modulo $m$).

Therefore, the $m$ equations~\eqref{eq:linearstress3} are linearly dependent.
They can be represented by $m-1$ equations: by breaking the cycle at an
arbitrary position and removing one edge, the simple cycle is transformed to a
path with a set of independent linear equations.
\hfill $\Box$

The implications of Theorem~1 are profound, namely: \\
{\em The steady-state stress in any structure with cycles can be solved by
removing edges to make it acyclic, yielding a spanning tree structure, which is
then solved to obtain the stress at all nodes.}

\subsection{Solving the steady-state analysis equations}
\label{sec:tree_solution}

\noindent
We will first analyze a tree structure, since, as shown above, the steady
state difference equations~\eqref{eq:linearstress2} are to be solved over a
spanning tree of a general interconnect structure.  

We choose an arbitrary leaf node of the tree as a reference; without loss of
generality, we will refer to it as node $v_1$, and the stress at that node as
$\sigma^1$.  For any node $v_i$ in the tree, there is a unique directed path
${\cal P}_i$ from $v_1$ to $v_i$, where each edge $e_k = (v_{s,k},v_{t,k}) \in
{\cal P}_i$ has a direction from $v_{s,k}$ to $v_{t,k}$ where $v_{s,k}$ is the
vertex that is closer to $v_1$.  Note that edges on this path are {\em
directed} from $v_1$ towards $v_i$.  However, it is built on an {\em
undirected} graph for the tree, where each undirected edge of the tree
has a reference current direction.

\begin{figure}
\centering
\includegraphics[width=0.7\linewidth]{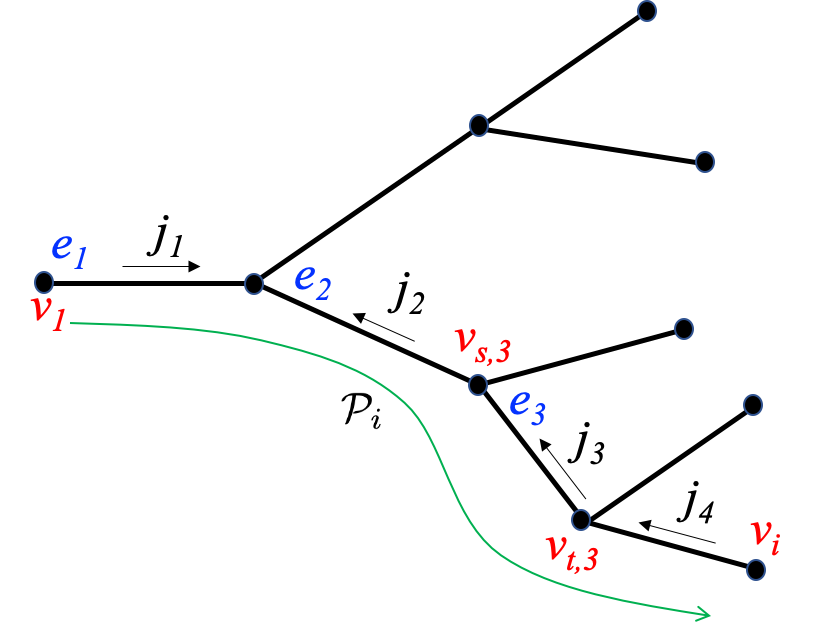}
\caption{An example undirected graph of a tree-structured interconnect, 
showing path ${\cal P}_i$ from reference node $v_1$ to node $v_i$.}
\label{fig:tree_ex}
\vspace{-4mm}
\end{figure}

To illustrate this point, consider the tree in Fig.~\ref{fig:tree_ex}, with 
path ${\cal P}_i$ from vertex $v_1$ to $v_i$. Vertex $v_{s,3}$ is the vertex of
$e_3$ that is closer to $v_1$.  The reference current directions on the
undirected graph are as shown: the direction of $j_1$ is along the direction of
path ${\cal P}_i$, while $j_2$, $j_3$, and $j_4$ are in the opposite direction.

\noindent
{\em Definition}:
We define $B_{{\cal P}_i}$, the ``Blech sum'' for a path ${\cal P}_i$, as:
\begin{equation}
B_{{\cal P}_i} = \sum_{e_k \in {\cal P}_i} \hat{j}_k l_k
\label{eq:Bi}
\end{equation}
where the summation is carried out over all edges $e_k$ on path ${\cal P}_i$.
The term $\hat{j}_k = j_k$ if the reference current direction for edge $e_k$
is in the same as path ${\cal P}_i$; otherwise, $\hat{j}_k = -j_k$.
Informally, $B_{{\cal P}_i}$ is the algebraic $(j l)$ sum along ${\cal P}_i$
from $v_1$ to $v_i$.

In the example of Fig.~\ref{fig:tree_ex}, the Blech sum to $v_{s,3}$ is
\begin{equation*}
B_{{\cal P}_{s,3}} = j_1 l_1 - j_2 l_2
\end{equation*}

\noindent
{\em Lemma~2}:
The stress, $\sigma^{i}$ at node $v_i$ is related to $\sigma^1$ as follows:
\begin{align}
\sigma^i = \sigma^1 - \beta B_{{\cal P}_i} \label{eq:sigma_i}
\end{align}

\noindent
{\em Proof:}
In a tree, the path ${\cal P}_i$ must be unique~\cite{Cormen09}.  Along this
path, the current on each edge $e_k$ from $v_{s,k}$ to $v_{t,k}$ is $\hat{j}_k$,
i.e., $j_k$ if the reference current direction is from $v_{s,k}$ to $v_{t,k}$,
and $-j_k$ otherwise.  Therefore, from~\eqref{eq:linearstress2}, 
\begin{eqnarray}
\sigma^{t,k} - \sigma^{s,k} = -\beta \hat{j}_k l_k
\label{eq:diffconstraints}
\end{eqnarray}
The continuity boundary condition~\eqref{eq:BC_continuity_tree} ensures that
the stress at the distal end of an edge on ${\cal P}_i$ is identical to that on
the proximal end of its succeeding edge, i.e., for successive edges $e_k$ and
$e_l$ on ${\cal P}_i$, $\sigma^{t,k} = \sigma^{s,l}$.  Therefore, adding these
equations over all edges on path ${\cal P}_k$, we see that as successive edges
on the path share a vertex $v$, $\sigma^v$ cancels out telescopically, except
for $v = v_1$ or $v_i$.  Meanwhile, the $\beta \hat{j}_k l_k$ terms add up, so
that the sum of all equations yields
\begin{align}
\sigma^i - \sigma^1 = - \beta 
               \sum_{e_k \in {\cal P}_i} \hat{j}_k l_k
\end{align}
This leads to the result in~\eqref{eq:sigma_i}. \hfill $\Box$

However,~\eqref{eq:sigma_i} in Lemma~2 stops short of determining $\sigma^i$ at
each node: for a tree with $|V|$ nodes, the lemma provides ($|V|-1$) linear
equations in $|V|$ variables, leading to an underdetermined system where each
node stress is related to the stress, $\sigma^1$, at an arbitrarily chosen leaf
node, $n_1$.  The $|V|^{\rm th}$ equation is obtained from the principle of
the conservation of mass: atoms are transported along a wire, but with zero net
change in the number of atoms in the wire.

\noindent
{\em Lemma~3}:
For a general tree/mesh interconnect with $|E|$ edges, with edge $k$
having width $w_k$ and height $h_k$,
\begin{align}
\sum_{k=1}^{|E|} w_k h_k \int_0^{l_k} \sigma_k(x) dx = 0
\label{eq:masscons}
\end{align}
The proof of the lemma is in the appendix and generalizes a similar result
from~\cite{Haznedar06}.  In effect, this is an integral form of the
BCs~\eqref{eq:BC_internal_tree_flux}, which conserve flux at the boundary of
each segment in the tree.  

\noindent
{\em Theorem 2}:
A tree or mesh interconnect with $|E|$ edges and $|V|$ vertices is immortal
when:
\begin{align}
\max_{1 \leq i \leq |V|} & \left ( \sigma^i \right ) < \sigma_{crit}
     \label{eq:max_vs_crit} \\
\mbox{where  }
\sigma^i &= 
\beta \left [ \frac{\sum_{k=1}^{|E|} w_k h_k
	\left [ \hat{j}_k \frac{l_k^2}{2} - B_{{\cal P}_{s,k}} l_k \right ]}
     {\sum_{k=1}^{|E|} w_k h_k l_k} - B_{{\cal P}_i} \right ]
     \label{eq:sigmaifinal}
\end{align}
where $B_{{\cal P}_i}$ is the ``Blech sum'' defined in~\eqref{eq:Bi}.

\noindent
{\em Proof:}
We first show that expression~\eqref{eq:sigmaifinal} provides the stress at
node $n_i$ of the interconnect, and is obtained by combining the result of
Lemma~3 with the $(|V|-1)$ equations from~\eqref{eq:sigma_i}.

Let edge $e_k$ connect vertices $v_{s,k}$ and $v_{t,k}$, where $v_{s,k}$ is the
vertex that is closer in the tree to the reference node $v_1$. Then,
substituting the result of Lemma~2 into Corollary~1,
\begin{align}
\int_0^{l_k} \sigma(x) dx =
\left ( \sigma^1 - \beta B_{{\cal P}_{s,k}} \right ) l_k
	- \beta \hat{j}_k \frac{l_k^2}{2}
\end{align}
where $B_{{\cal P}_{s,k}}$ is the Blech sum from node $n_1$ to node
$v_{s,k}$.\footnote{The use of $\hat{j}_k$ allows for the traversal
from $v_1$ to $v_i$ to include edges in a direction opposite to the reference
current direction: the stress difference between nodes on such edges should
have the opposite sign as~\eqref{eq:linearstress2} in Lemma~1.}

Substituting the integral expressions in~\eqref{eq:masscons} from Lemma~3:
\begin{align}
\sum_{k=1}^{|E|} w_k h_k \left [
\left ( \sigma^1 - \beta B_{{\cal P}_{s,k}} \right ) l_k
	- \beta \hat{j}_k \frac{l_k^2}{2} \right ] = 0
\label{eq:masscons2}
\end{align}
After further algebraic manipulations, we obtain
\begin{align}
\sigma^1 =
  \frac{\beta \sum_{k=1}^{|E|} w_k h_k
	\left [ \hat{j}_k \frac{l_k^2}{2} + B_{{\cal P}_{s,k}} l_k \right ]}
     {\sum_{k=1}^{|E|} w_k h_k l_k}
\label{eq:sigma1}
\end{align}
Finally, we substitute the above into~\eqref{eq:sigma_i} to
obtain~\eqref{eq:sigmaifinal}, the expression for the steady-state stress
values at each node $i$.

For the interconnect to be immortal, the largest value of stress in the tree
must be lower than $\sigma_{crit}$, the critical stress required to induce a
void.  From Corollary 2, in finding the maximum stress in the tree, it is
sufficient to examine the stress at the nodes of the tree, so that the largest
node stress is below $\sigma_{crit}$. This proves~\eqref{eq:max_vs_crit}.
\hfill $\Box$

\section{Linear-Time Immortality Calculation}
\label{sec:solution}

\noindent
As we have established, a general interconnect on a graph can be solved by
considering the solution of Theorem~1 on a tree of the graph.  Identifying such
tree is straightforward, and standard methods such as depth-first or
breadth-first traversal can be used.

After arriving at a tree structure, although Theorem~2 provides a useful,
closed-form result, a simple-minded computation would calculate $\sigma^i$ at
each node $v_i$ in the tree through repeated incantations
of~\eqref{eq:sigmaifinal}.  However, as we will show, this computation can be
performed in $O(|E|)$ time for a structure with $|E|$ edges.  We
rewrite~\eqref{eq:sigmaifinal} as:
\begin{align}
\sigma^i &= \beta \left [ \frac{Q}{A} - B_{{\cal P}_i} \right ]
\label{eq:sigmai_rewritten} \\
\mbox{where   } 
Q &= \textstyle \sum_{k=1}^{|E|} w_k h_k \left [
           \hat{j}_k \frac{l_k^2}{2} + B_{{\cal P}_{s,k}} l_k \right ] 
\label{eq:Qcomp} \\
A &= \textstyle \sum_{k=1}^{|E|} w_k h_k l_k 
\label{eq:Acomp}
\end{align}
This computation requires the calculation of three summations for 
$A$, $Q$, and for the Blech sum, $B_{{\cal P}_i}$ from reference node $v_1$ to
each node $i$ in the tree.  It proceeds in the following steps:
\begin{enumerate}
\item[1.]
To compute $B_{{\cal P}_i}$, we traverse the tree from $v_1$ using a standard
traversal method, e.g., the breadth-first search (BFS). At node $v_1$, we
initialize $B_{{\cal P}_{v_1}} = 0$. As we traverse each edge $e_k =
(v_{s,k},v_{t,k})$, we compute $B_{{\cal P}_{t,k}}$.  
\item[2.]
Using the above Blech sums to each node, we compute $Q$ (Eq.~\eqref{eq:Qcomp})
and $A$ (Eq.~\eqref{eq:Acomp}), summing over all edges.
\item[3.]
Finally, we compute $\sigma^i$ at each node $i$
using~\eqref{eq:sigmai_rewritten}.
\end{enumerate}
{\bf Complexity analysis}: The BFS traversal in Step~1 over a tree traverses
$O(|E|)$ edges. For each edge, Step~2 performs a constant number of
computations to obtain $A$ and $Q$ (~\eqref{eq:Acomp}--\eqref{eq:Qcomp}). The
final computation of~\eqref{eq:sigmai_rewritten} in Step~3, and the immortality
check that compares the computed value with $(\sigma_{crit}-\sigma_T)$
according to~\eqref{eq:max_vs_crit}, perform a constant number of computations
for $|V|$ nodes.  Therefore, the computational complexity for any tree
structure is $O(|E|)$.

\begin{figure}[htb]
\centering
\vspace*{-4mm}
\includegraphics[width=0.5\linewidth]{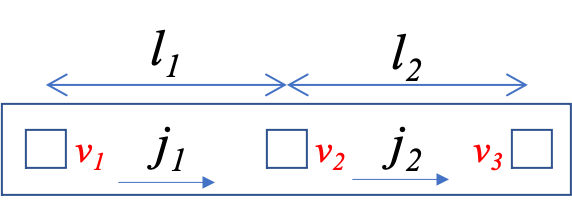}
\caption{A two-segment interconnect line.}
\label{fig:twoseg}
\vspace*{-2mm}
\end{figure}

\noindent
{\bf Example}:
We illustrate our computation for a two-segment line (Fig.~\ref{fig:twoseg}) in
a single layer (with constant $h_k$) in Table~\ref{tbl:twoseg}, using the
leftmost node $v_1$ as the reference.  Starting from $v_1$, the two edges are
traversed to compute $B$.  The symbol $B_{{\cal P}_{t,k}}$ represents the Blech
sum calculated at the distal vertex $v_{t,k}$ of the edge; note that the
computation of $Q$ uses the Blech sum at the proximal vertex, $v_{s,k}$.

\begin{table}[hbtp]
\centering
\caption{Sequence of computations for a two-segment wire.}
\begin{tabular}{|l|c|c|c|}
\hline
                 & $A$                 & $B_{{\cal P}_{t,k}}$ & $Q$ \\ \hline \hline
Initializaton    & 0                   & 0                    & 0 \\ \hline
Edge $(v_1,v_2)$ & $w_1 l_1$           & $j_1 l_1$            & $w_1 j_1 l_1^2/2$
\\ \hline
\multirow{2}{*}{Edge $(v_2,v_3)$}
                 & \multirow{2}{*}{$w_1 l_1 + w_2 l_2$}
                                       & \multirow{2}{*}{$j_1 l_1 + j_2 l_2$}
                                                              & $w_1 j_1 l_1^2/2 + 
                                                                  w_2 j_2 l_2^2/2$ 
\\
                 &                     &                      &  $+ w_2 l_2 (j_1 l_1)$
\\ \hline
\end{tabular}
\label{tbl:twoseg}
\end{table}

Based on the table, we compute the stress at each node as:
\begin{align}
\sigma^{v_1} = \beta \frac{w_1 j_1 l_1^2 + w_2 j_2 l_2^2 + 2 w_2 j_1 l_1 l_2}
                          {2(w_1 l_1 + w_2 l_2)} \label{eq:ss_line_stress} \\
\sigma^{v_2} = \sigma^{v_1} - \beta (j_1 l_1)
\; \; ; \; \;
\sigma^{v_3} = \sigma^{v_1} - \beta (j_1 l_1 + j_2 l_2) \nonumber 
\end{align}
The analysis of this line in~\cite{Sun18} yields an identical result;
unlike our method, \cite{Sun18} cannot analyze arbitrary trees/meshes in
linear time.

\begin{figure*}
\resizebox{1.05\linewidth}{!}{%
\centering
\hspace*{-8mm}
\includegraphics[width=0.33\linewidth]{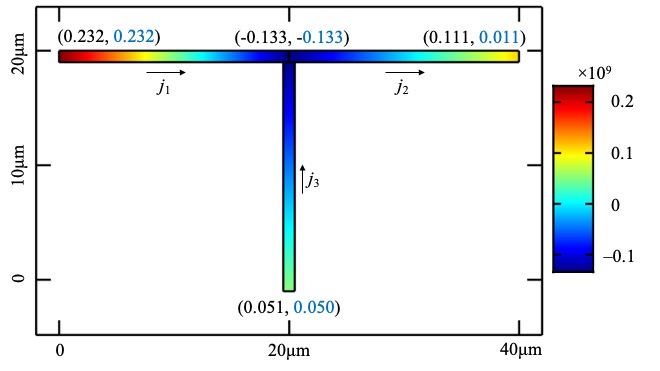}
\hspace*{-3mm}
\includegraphics[width=0.33\linewidth]{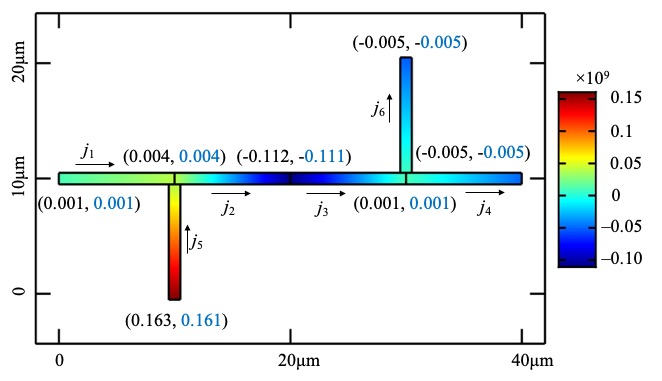}
\hspace*{-3mm}
\includegraphics[width=0.31\linewidth]{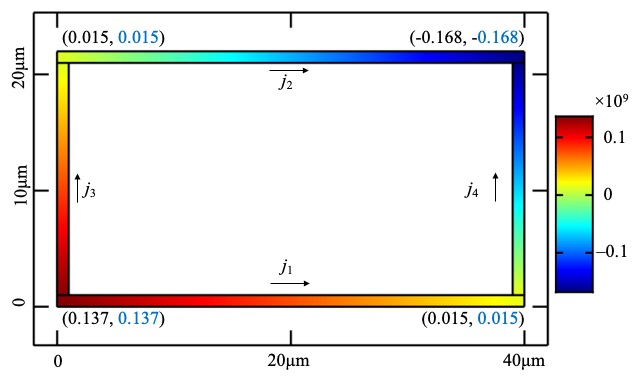}
}
\caption{Comparison of the steady-state stress in three structures: a T,
a tree, and a mesh.  The stress at each node (in GPa) is shown as a tuple, with
our closed-form solution in black and the COMSOL solution in blue text.  The
color bar is based on a COMSOL solution.  The width of each segment
is $1\mu$m, and length scales are shown in the figure. The current densities
in the T are $j_1  = 6 \times 10^{10} A/m^2 $ , $j_2 =
-4  \times 10^{10}$A/m$^2$ , $j_3 = 3\times 10^{10}$A/m$^2$. 
For the tree, $j_1 = -1 \times 10^{10}$A/m$^2$, $j_2 = 5 \times 10^{10}$A/m$^2$,
$j_3 = -4\times 10^{10}$A/m$^2$, $j_4 = j_6 = 2 \times
10^{10}$A/m$^2$, $j_5 = 4 \times 10^{10}$A/m$^2$.  For the mesh structure, $j_1 =
1\times 10^{10}$A/m$^2$, $j_2 = 1.5\times 10^{10}$A/m$^2$, $j_3 =  2\times 10^{10}$A/m$^2$,
$j_4 = 3\times 10^{10}$A/m$^2$.}

\label{fig:COMSOL_color_map}
\end{figure*}

\section{Results}
\label{sec:results}

\noindent
We present three sets of results.  The first set shows comparisons
with a numerical solver in Section~\ref{sec:COMSOL}.  Next, in
Section~\ref{sec:IBM}, we use our method to analyze large public-domain IBM
power grid benchmarks.  These were designed for old Al lines, but we
assume them to be modern Cu DD wires.  The large sizes of these benchmarks test the
scalability of our approach.  Finally, in Section~\ref{sec:OpeNPDN}, we perform
analysis of power grids on designs synthesized on a commercial 28nm and
Nangate 45nm parameters, both based on Cu DD interconnects.

In Cu DD based technologies, each layer can be treated separately due
to the presence of barrier/capping layers that prevent atomic flux from flowing
across layers through vias.  The methods in this paper are applied to each
layer to find the steady-state stress, which is then used to predict
immortality.  This limits the size of the EM problem, since it must be solved
in a single layer at a time.  Moreover, since it is common to use a reserved
layer model where all wires in a layer are in the same direction, effectively
this implies that each layer consists of a set of metal lines with a limited
number of nodes.  In such scenarios, the EM problem reduces to the analysis of
a large number of line/tree structures, each with tens of nodes.  The IBM
benchmarks contain mesh structures within layers, which enable us to better
evaluate our method.

\subsection{Comparison with COMSOL}
\label{sec:COMSOL}

\noindent
We show comparisons between our approach and numerical simulations using
COMSOL on Cu DD based structures.  The material parameters, provided to COMSOL,
are~\cite{ala:05}:
$\rho =$ 2.25e-8$\Omega$m,
${\cal B} =$ 28GPa,
$\Omega =$ 1.18e-29m$^3$,
$D_0 =$ 1.3e-9m$^2$/s,
$E_a =$ 0.8eV,
$Z^* =$ 1,
$\sigma_{crit} = 41$MPa,
$T = 378$K.
COMSOL is limited to analyzing small structures, which is reflected 
the topologies shown in Fig.~\ref{fig:COMSOL_color_map}:
\begin{itemize}
\item
An interconnect tree with three segments
\item
A larger interconnect tree
\item
A simple mesh structure
\end{itemize}

The color maps in the figure show the spatial variation of steady-state stress
over each interconnect, where the numbers next to each node represent the
values computed using our approach and by COMSOL. It is easily seen that the
numbers match; since our approach is exact, any discrepancies can be attributed
to numerical inaccuracies in COMSOL.

\subsection{Analysis on IBM power grid benchmarks}
\label{sec:IBM}

\noindent
The only widely available power grid benchmarks are the IBM
benchmarks~\cite{IBMPDN_url}.  Each benchmark contains Vdd and Vss networks and
multiple voltage domains, and general tree/mesh structures in individual
layers.  We implement a BFS traversal over these structures using Python3.6
and Deep Graph Library~\cite{karypis20} on a GPU by modifying the message
passing functions. Run times are shown on a 3.6GHz
Intel 
Core 
i7-7820X and NVIDIA RTX 2080Ti GPU.  

\begin{figure}[htb]
\centering
\includegraphics[width=0.63\linewidth]{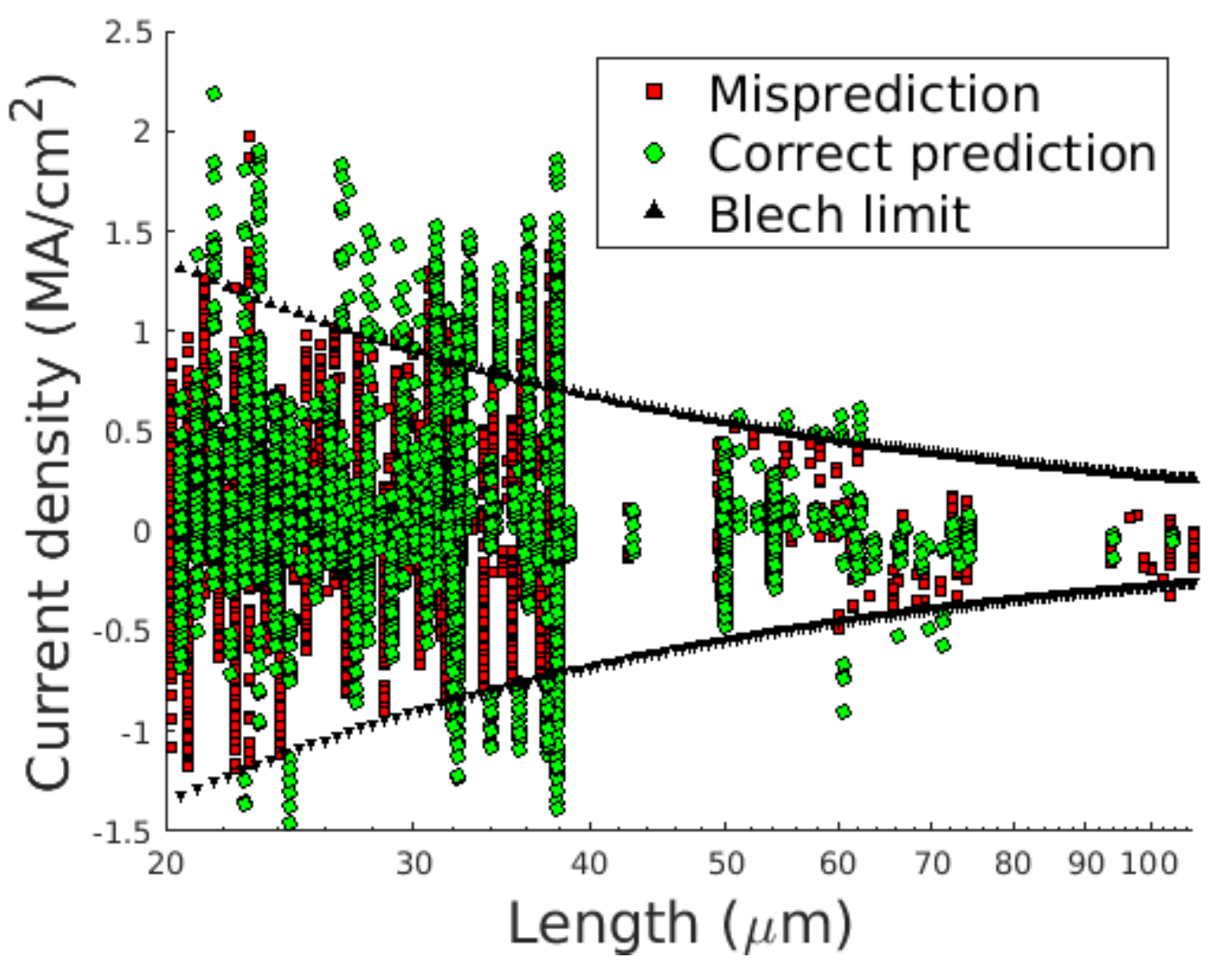}
\caption{Inaccuracy of the traditional Blech filter (ibmpg6).}
\label{fig:scatter-ibmpg6}
\end{figure}

The traditional Blech criterion is only accurate for a single-segment wire:
next, we evaluate its accuracy.  We consider our approach as the accurate
result since it is rigorously derived for multisegment structures by
generalizing the same physics-based modeling framework used by the Blech
criterion for one-segment wires, and it is validated on COMSOL.  Therefore a
positive identification of immortality implies that our method
finds the segment to be immortal; a negative identification implies mortality.

Fig.~\ref{fig:scatter-ibmpg6} plots the current density $j$ vs. the wire length
$l$ within the segments of the ibmpg6 benchmark.  The currents in the Vdd and
Vss lines may be either positive or negative, and their magnitude affects EM.
The black triangles show the contours of $jl = (jl)_{crit}$: when the magnitude
lies within this frontier for a segment of the grid, the traditional Blech
criterion~\eqref{eqn:Blech_criterion} would label the wire as immortal;
otherwise it is potentially mortal.  To help highlight erroneous predictions,
the figure shows green markers for correct predictions and red markers for
incorrect predictions. The Blech criterion shows significant inaccuracy 
on multisegment wires.

\begin{table}
\centering
\caption{Comparison of our approach against the traditional Blech filter on the IBM benchmarks (TP = true positive, TN = true negative, FP = false positive, FN = false negative.)}
\label{tbl:table-ibm}
\resizebox{\linewidth}{!}{%
\begin{tabular}{||c||r||r|r|r|r||r|r||} 
\hhline{|t:=:t:=:t:====:t:==|}
 & \multicolumn{1}{c||}{$|E|$} & \multicolumn{1}{c|}{TP} & \multicolumn{1}{c|}{TN} & \multicolumn{1}{c|}{{\bf FP}} & \multicolumn{1}{c||}{\bf{FN}} & \multicolumn{2}{c|}{Runtime} \\ 
\cline{7-8}
 &     &        &           &              &                  & GPU  & CPU  \\
\hhline{|t:=:t:=:t:====:t:==|}
pg1 & 29750 & 1557 & 10144 & \textbf{17372} & \textbf{677} & 7s & 6s \\ 
\hline
pg2 & 125668 & 7703 & 33534 & \textbf{82025} & \textbf{2406} & 12s & 19s \\ 
\hline
pg3 & 835071 & 200158 & 3539 & \textbf{630979} & \textbf{395} & 36s & 184s \\ 
\hline
pg6 & 1648621 & 916094 & 1365 & \textbf{730995} & \textbf{167} & 88s & 280s \\
\hhline{|b:=b:=:b:====:b:==|}
\end{tabular}
}
\vspace{-4mm}
\end{table}

Table~\ref{tbl:table-ibm} summarizes the results on IBM benchmarks.  
True positives (TP) and true negatives (TN) correspond to correct predictions
where the Blech criterion agrees with our accurate analysis.  The errors
correspond to {\em false negatives} (FN), where an immortal segment is deemed
potentially mortal by the traditional Blech criterion, and {\em false
positives} (FP), where an mortal segment is labeled as potentially immortal by
Blech. FPs cause failures to be overlooked, and FNs may lead to
overdesign as EM-immortal wires are needlessly optimized.

The table shows that:
\begin{itemize}
\item
the inaccuracies in the Blech filter are seen across benchmarks.
\item
our method is scalable to large mesh sizes with low runtimes.
\end{itemize}

From the data, it is apparent that the traditional Blech criterion can provide
misleading results. The reasons for this are twofold:
\begin{itemize}
\item
A high-$jl$ segment could be immortal if it has numerous downstream segments
with low $jl$, so that the total $jl$ sum may be low. For example, in Fig.~\ref{fig:twoseg}, if the current density $j_1 = 0$, then the segment acts as passive
reservoir, bringing down the stress in the right segment to be lower than
the case of an identical isolated segment carrying the same current, but with
a blocking boundary at $v_2$~\cite{Lin16}.  
\item
A low-$jl$ segment could be labeled immortal by the traditional criterion,
but it may be mortal due to a high stress at one node, caused by a high Blech
sum for downstream wire segments, which could raise the stress at the other
node.
\end{itemize}

\subsection{Analysis on OpenROAD power grids}
\label{sec:OpeNPDN}

\noindent
We show simulations based on power grids from circuits designed using a
commercial 28nm and Nangate45 technologies using Cu DD interconnects.  The
circuits are taken through synthesis, placement and routing in these technology nodes
(some circuits are implemented in both nodes) using a standard design flow.
The power grid is synthesized using an open-source tool,
OpeNPDN~\cite{Chhabria20OpeNPDN} from OpenROAD.  The IR drop and currents are
computed using PDNSim~\cite{PDNSim}, with currents scaled to provide an IR drop
of 5mV.

\begin{figure}[htb]
\centering
\vspace{-2mm}
\includegraphics[width=0.65\linewidth]{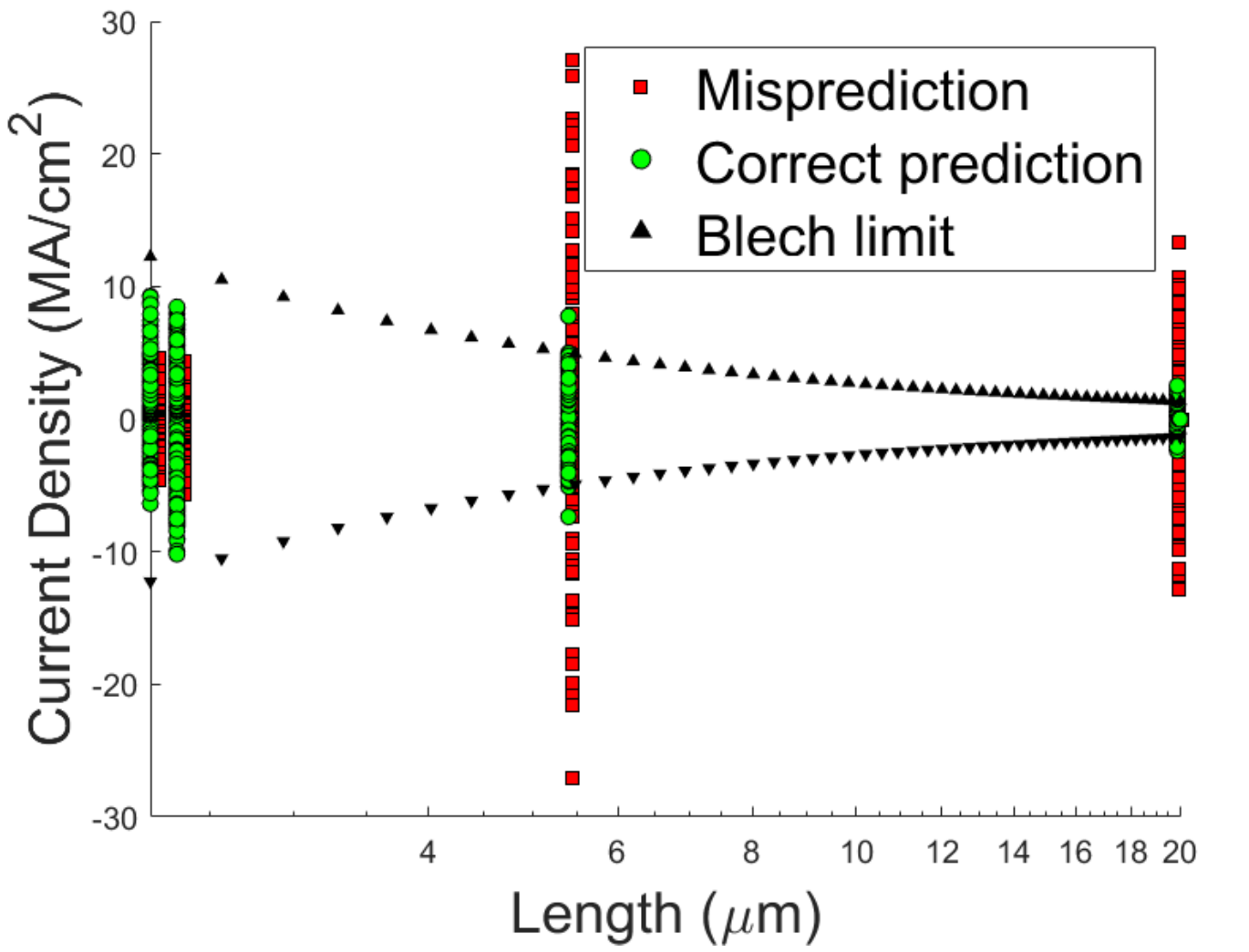}
\caption{Inaccuracy of traditional Blech filter (jpeg/28nm).}
\label{fig:scatter-OpeNPDN}
\vspace{-4mm}
\end{figure}

Fig.~\ref{fig:scatter-OpeNPDN} shows a scatter plot that analyzes the
inaccuracy of the traditional Blech criterion on a Cu-based technology, using
$(jl)_{crit} = 0.27$A/$\mu$m, based on material parameters listed in
Section~\ref{sec:COMSOL}. Due to the regular structure of the power grid, 
many lines have the same length. As in the earlier case, it is easily seen that
the Blech criterion leads to numerous false positives and false negatives.
Results for more circuits are listed in Table~\ref{tbl:OpeNPDN}, and show
similar trends.

\begin{table}
\centering
\caption{Comparison of our approach against the traditional Blech filter on a 28nm technology with Cu interconnects.}
\label{tbl:OpeNPDN}
\resizebox{\linewidth}{!}{%
\begin{tabular}{||c||l||r||r|r|r|r||} 
\hhline{|t:=t:=:t:=:t:====|}
& Circuit & \multicolumn{1}{c||}{$|E|$} & TP & TN & {\bf FP} & {\bf FN} \\
\hhline{|:=::=::=::====|}
\multirow{3}{*}{28nm} 
& gcd   & 678        & 634   &8       & {\bf 31}     & {\bf 5}      \\
\cline{2-7}
& aes   & 11,361     & 8,039 &0       & {\bf 3,297}  & {\bf 25}     \\
\cline{2-7}
& jpeg  & 123,220    & 63,889 & 71    & {\bf 58,696} & {\bf 564}    \\
\hhline{|:=:=::=::====|}
\multirow{4}{*}{45nm} 
& dynamic\_node      
        & 6,270      & 2,617  & 256   & {\bf 3,059}  & {\bf 338}    \\
\cline{2-7}
& aes   & 7,212      & 3,255  & 322   & {\bf 3,160}  & {\bf 475}    \\
\cline{2-7}
& ibex  & 12,128     & 4,645  & 1,112 & {\bf 4,964}  & {\bf 1,407}  \\
\cline{2-7}
& jpeg  & 35,848     & 10,052 & 5,047 & {\bf 15,479} & {\bf 5,270}  \\
\cline{2-7}
& swerv & 59,049     & 14,545 & 9,762 & {\bf 23,366} & {\bf 11,376} \\
\hhline{|b:=b:=:b:=:b:====|}
\end{tabular}
}
\vspace{-3mm}
\end{table}

\ignore{
\subsection{\redHL{Why are tree structures enough?}}

\noindent
In deeply-scaled FinFET technologies using Cu DD interconnects,
it is adequate to perform EM analysis on tree structures:
\begin{enumerate}
\item As stated above, all mass transfer occurs within each metal layer as
migrating atoms are prevented from moving across vias across metal layers due
to blocking boundaries in Cu DD interconnects~\cite{Gambino18,Zhang10}.
\item Lithography considerations dictate that especially in lower metal layers
(which are seen to be susceptible to electromigration~\cite{Pande19}), wires must
be routed unidirectionally.  As a result, all wires in a given layer are
oriented in the same direction, and line structures are the most commonly
encountered structures.
\item In principle, it is possible to use bidirectional routing and mesh structure
in upper metal layers that are less constrained by lithography. However, design
methodologies widely use unidrectional routing and create meshes by connecting
grids of parallel orthogonal wires in successive
layers~\cite{Chan02,Chang14,Chhabria20}.  As a result, in each layer, separated
from neighboring layers by blocking boundaries, the resulting structure is a
line or a tree.
\end{enumerate}
Therefore, for modern chips, it is sufficient to consider the analysis of lines
and tree structures rathern than meshes.
}

\section{Conclusion}

\noindent
A linear-time approach for checking immortality in a general tree/mesh
interconnect is proposed. The results are validated against COMSOL and
shown to be fast and scalable to large power grids.

\section*{Appendix: Proof of Lemma 3}

\noindent
{\em Proof:}
The stress on a wire segment causes a displacement of $u_i$ in segment $i$ of
the interconnect structure.  The stress has no shear component since the
current in a line is unidirectional.  Due to conservation of mass, the net
material coming from all $|E|$ wire segments is zero, and therefore,
\begin{align}
\textstyle \sum_{k=1}^{|E|} w_k h_k u_k = 0
\label{eq:massconv}
\end{align}
where $w_k$ is the width of the $k^{\rm th}$ wire segment.  The displacement
$u_k$ is the integral of displacements $du_k$ over the segment caused by
stress $\sigma_k(x)$ applied on elements of size $dx$ in segment $k$.  If
${\cal B}$ is the bulk modulus, from Hooke's law,
\begin{align}
u_k = \textstyle \int_0^{l_k} du_k(x) 
    = {\cal B} \textstyle \int_0^{l_k} \sigma_k(x) dx
\end{align}
Combining this with~\eqref{eq:massconv} leads to the result of Lemma~3.
\hfill $\Box$

\bibliographystyle{misc/ieeetr2}
\bibliography{bib/main,bib/main3}
\end{document}